\documentclass[sigconf]{acmart}

\AtBeginDocument{%
  }

\setcopyright{none}
\acmDOI{}
\acmISBN{}
\acmConference[CHI 2026 Workshop]{CHI 2026 Workshop}{2026}{}
\acmYear{2026}
\copyrightyear{2026}

\renewcommand\footnotetextcopyrightpermission[1]{}

\title[Circadian Phase Locking of Epilepsy Seizures in Wearable Data]{{Circadian Phase Locking of Epilepsy Seizures in Wearable Data: A Single-Patient Case Study}}

\author{Berenika Ewart-James}
\affiliation{%
  \institution{University of Bristol}
  \city{Bristol}
  \country{United Kingdom}}
\email{berenika.ewart-james@bristol.ac.uk}

\author{Matthew Wragg}
\affiliation{%
  \institution{University of Bristol}
  \city{Bristol}
  \country{United Kingdom}}
\email{matthew.wragg@bristol.ac.uk}
  
\author{Nawid Keshtmand}
\affiliation{%
  \institution{University of Bristol}
  \city{Bristol}
  \country{United Kingdom}}
\email{nawid.keshtmand@bristol.ac.uk}
  
\author{Amberly Brigden}
\affiliation{%
  \institution{University of Bristol}
  \city{Bristol}
  \country{United Kingdom}}
 \email{amberly.brigden@bristol.ac.uk}
 
\author{Paul Marshall}
\affiliation{%
  \institution{University of Bristol}
  \city{Bristol}
  \country{United Kingdom}}
 \email{p.marshall@bristol.ac.uk}

\author{Raul Santos-Rodriguez}
\affiliation{%
  \institution{University of Bristol}
  \city{Bristol}
  \country{United Kingdom}}
 \email{enrsr@bristol.ac.uk}

\begin{abstract}

Epilepsy is a common, chronic neurological disorder characterized by recurrent seizures caused by sudden bursts of abnormal electrical activity in the brain. Seizures can often be unpredictable, leading to uncertainty and anxiety for people with epilepsy (PWE). To address this problem, the Epilepsy UK Priority Setting Partnership (PSP, n=2798) identified research into seizure forecasting technology as a priority.
Seizure onsets are recorded as discrete events embedded within continuously sampled physiological signals that exhibit strong circadian and multi-day rhythms. Standard modelling approaches often treat time as linear or rely on clock-time features, which may not explicitly capture the underlying physiological phase. In this paper, we examine whether seizure onsets exhibit phase preference relative to circadian rhythms derived from wearable inter-beat interval (IBI) data. As a proof-of-concept, using 176 days of wearable and seizure diary data from a single patient, we extract oscillatory components via band-limited filtering and Hilbert-based phase estimation, and test for non-uniform seizure–phase alignment using circular statistics. We observe significant circadian phase concentration ($R=0.55$, Rayleigh $p=7.5\times10^{-5}$), while multi-day bands do not show consistent or statistically significant phase clustering in this dataset. Exploratory logistic baselines indicate modest but detectable structure beyond simple clock-time effects. We argue that explicit physiological phase representations provide an interpretable bridge between continuous wearable sensing and sparse clinical events, and may augment existing seizure forecasting pipelines. We discuss implications for multi-scale modelling, patient-facing interfaces, and future multi-patient validation.

\end{abstract}



\begin{document}
\maketitle


\section{Introduction and Related Work}

Epilepsy is characterised by recurrent seizures that are often experienced as unpredictable. This perceived unpredictability imposes substantial psychological and practical burden, motivating efforts to develop seizure forecasting systems that support safer daily living and improved self-management. The ATMOSPHERE project (Artificial intelligence To Optimise Seizure Prediction to Empower people with Epilepsy) \cite{quilter_digital_2024} is exploring seizure forecasting technology combining wearable and smartphone sensing with machine learning algorithms to enable real-time, individualised seizure risk estimation.

Everyday wearable devices such as smartwatches generate dense, longitudinal physiological streams, including heart rate, inter-beat interval (IBI), activity levels, and sleep timing. These signals are continuously sampled and exhibit well-established circadian structure, and in some cases multi-day oscillations \cite{baud_multi-day_2018,karoly_multiday_2021}. In contrast, seizure onsets are recorded as discrete events within these continuous streams. For many ambulatory epilepsy datasets, seizure occurrences are relatively infrequent compared to the high temporal resolution of wearable sensing. This representational mismatch presents a modelling challenge: many statistical and machine-learning approaches treat observations as independent or encode time using linear or clock-based features, whereas seizure risk may fluctuate according to underlying biological rhythms \cite{baud_multi-day_2018,gregg_seizure_2023}. Evidence from chronic brain recordings demonstrates circadian and multi-day modulation of seizure likelihood, and peripheral signals such as heart rate have been shown to exhibit multi-day rhythmic structure associated with seizure clustering \cite{karoly_multiday_2021}.

Despite progress in seizure prediction using wearable devices, significant challenges are still observable. Wearable signals are noisy, missingness is common, behavioural and medication effects confound interpretation, and seizure labels form the intended user groups are sparse. Prior studies combining heart rate, sleep, and activity features with self-reported seizures demonstrate the technical feasibility of seizure forecasting using wearable-derived signals, but report heterogeneous predictive performance and limited model interpretability \cite{stirling_forecasting_2021,mason_heart_2024}. Importantly, most existing forecasting approaches treat time as a linear variable or encode it using simple clock-time features (e.g., hour of day), rather than explicitly modelling where an individual lies within an underlying biological rhythm. As a result, cyclical modulation of seizure risk may be captured only indirectly, limiting both physiological interpretability and the ability to communicate rhythm-based vulnerability patterns to users. Making biological phase an explicit modelling variable may therefore support more transparent, human-centred representations of fluctuating seizure risk.

In this paper, we explore a cycle-aware modelling perspective that augments existing forecasting pipelines rather than replacing them. Conceptually, we ask whether seizures tend to occur at particular times within an individual’s daily biological rhythm, rather than being evenly distributed across the day. Using a conservative single-patient analysis, we examine whether seizure onsets cluster at specific phases of a circadian rhythm derived from wearable heart activity data (inter-beat interval, IBI), extracted using band-limited filtering and Hilbert-based phase estimation.
We observe significant circadian phase concentration in one case study, whereas multi-day rhythms did not show statistically significant or consistent phase clustering in this dataset. We argue that representing physiological phase explicitly provides an interpretable bridge between continuous wearable sensing and discrete clinical events, and may offer a principled foundation for future multi-scale and causal modelling in everyday wearable health systems.

\section{Methodology}

\subsection{Problem Formulation}

Let $t$ index discrete time points across the observation window. Let $X_t \in \mathbb{R}$ denote the wearable-derived physiological signal (e.g., inter-beat interval) at time $t$. Let $E = \{e_1, \dots, e_n\}$ denote the set of recorded seizure onset times within the same interval. Wearable signals are densely sampled and exhibit oscillatory structure, whereas seizure onsets form a sparse point process over the same observation window.

If seizure risk is modulated by underlying biological rhythms \cite{baud_multi-day_2018,karoly_multiday_2021,gregg_seizure_2023}, then seizure onsets should occur preferentially at specific phases of these oscillations rather than uniformly over time. We therefore test whether seizure onsets are non-uniformly distributed over oscillatory phase extracted from $X_t$.

If seizure risk is modulated by underlying biological rhythms \cite{baud_multi-day_2018,karoly_multiday_2021,gregg_seizure_2023}, then seizure onsets should occur preferentially at specific phases of these oscillations rather than uniformly over time. We therefore test whether seizure onsets are non-uniformly distributed over oscillatory phase extracted from $X_t$. In this context, we use the term “phase locking” to describe the tendency for seizure onsets to cluster at particular points within a repeating physiological rhythm rather than being evenly distributed across the cycle.

Establishing such phase preference is a necessary first step toward cycle-aware risk modelling and toward investigating cyclical causal structure in wearable-derived physiological data. This single-patient proof-of-concept prioritises interpretability and conservative signal processing over complex predictive modelling.

\subsection{Data and Oscillatory Signal Processing}

\begin{figure}[t]
  \centering
  \includegraphics[width=\columnwidth]{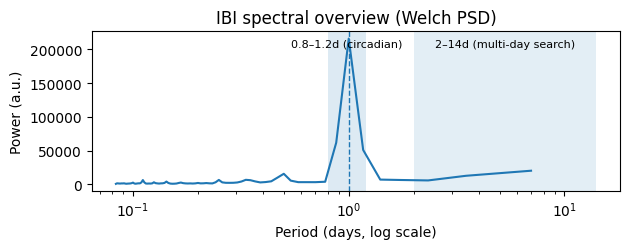}
  \caption{Spectral overview of hourly inter-beat interval (IBI) using Welch power spectral density (power vs period, log scale). A pronounced peak near 1 day confirms strong circadian rhythmicity in this dataset. In contrast, power across multi-day periods (2--14 days) is weaker and broadly distributed, without sharply defined peaks. Shaded regions indicate the circadian band selected for phase extraction (0.8--1.2 days) and the multi-day search range screened for exploratory phase locking. The clear dominance of the circadian component motivated prioritisation of circadian phase in subsequent analyses.}
  \label{fig:ibi_psd}
\end{figure}
We analysed wearable and seizure diary data from a single adult male patient (age range 35–44 years) with a clinical diagnosis of focal epilepsy (focal aware and focal impaired awareness seizures), recorded over 176 consecutive days. A total of 29 seizure onsets were reported during the observation period.
We analysed continuous wearable-derived inter-beat interval (IBI) data and nightly sleep summary scores from a single patient over 176 days.

IBI was resampled to a regular hourly grid ($\Delta t = 60$ minutes) to support spectral and phase-based analyses at circadian and multi-day time-scales while maintaining robustness to missingness and reducing high-frequency noise. An hourly resolution preserves the 24-hour rhythm (24 samples per day) and provides sufficient temporal granularity for mapping seizure onsets, while avoiding the instability and interpolation burden that can arise at finer sampling intervals. Sleep summary variables were represented at daily resolution using one value per night. Missing values were imputed using time-aware linear interpolation only for short gaps (IBI: $\leq 6$ consecutive hours; sleep: $\leq 2$ consecutive days). Longer gaps were retained as missing and excluded from subsequent spectral and phase analyses to reduce the risk of introducing artificial low-frequency structure.

Nightly sleep scores were assigned to the subsequent calendar day, reflecting the assumption that sleep influences next-day physiological state \cite{karoly_multiday_2021}. Z-score normalisation used a single mean and standard deviation per signal computed across the full study period (excluding missing values). Seizure diary events ($n=29$) were preserved as precise onset timestamps for phase mapping.

To identify dominant rhythmic components, Welch power spectral density (PSD) \cite{welch_use_1967} was computed for hourly IBI and daily sleep signals (Figure~\ref{fig:ibi_psd}). IBI exhibited a pronounced peak near 1 day, consistent with circadian rhythmicity, while multi-day power was weaker and broadly distributed.

Based on both spectral structure and prior literature, we defined a circadian band (0.8--1.2 days) around the 24-hour peak and screened exploratory multi-day bands spanning 2--28 days. In this dataset, circadian oscillations were the most clearly defined component; subsequent phase analyses therefore prioritised the circadian band, with multi-day bands treated as secondary analyses.

To isolate oscillatory components, we applied bandpass Butterworth filtering \cite{butterworth1930theory} using zero-phase forward--backward implementation to avoid phase distortion. Filtering was performed only on contiguous non-missing segments. Segments shorter than three cycles of the slowest band component were excluded to stabilise phase estimation and mitigate edge effects.

For each retained band-limited segment, we computed the analytic signal via the Hilbert transform \cite{boashash_estimating_1992} and extracted instantaneous phase $\phi(t)=\arg(z(t))$ and amplitude $A(t)=|z(t)|$, with $\phi(t)\in[-\pi,\pi]$. The Hilbert transform provides a continuous time-resolved phase estimate for narrowband oscillations without requiring explicit peak or trough detection, enabling direct mapping of seizure onset timestamps onto physiological phase.

\subsection{Phase Mapping and Statistical Analysis}

Seizure onsets were mapped to the closest available phase estimate within the band-limited time series. Events without defined phase (e.g., due to filtering boundaries) would be excluded; however, in this dataset no seizures were lost.

Phase concentration was quantified using the resultant vector length $R$ and statistical significance assessed using the Rayleigh test for non-uniformity \cite{fisher_statistical_1995}, appropriate for detecting unimodal clustering in circular data.

To contextualise phase effects relative to clock time, exploratory logistic models were fitted using (i) time-of-day, (ii) circadian phase, and (iii) nightly sleep score as individual predictors. Discriminative performance was evaluated using area under the receiver operating characteristic curve (AUC). These models are not intended as forecasting systems but provide simple baselines to assess whether oscillatory phase captures structure beyond clock-time or coarse sleep effects.

\section{Single-Patient Case Study}

\subsection{Spectral Screening}

As shown in Figure~\ref{fig:ibi_psd}, hourly IBI exhibited a clear and dominant spectral peak near 1 day, consistent with circadian rhythmicity. In contrast, multi-day power (2--28 days) was weaker and more broadly distributed, without sharply defined peaks.

This spectral profile motivated prioritisation of the circadian band for subsequent phase-based analyses, while retaining multi-day bands for exploratory screening.
\subsection{Phase Locking Results}

\begin{table}[t]
\caption{Phase-locking results across candidate IBI bands (single patient).}
\label{tab:band_scan}
\centering
\begin{tabular}{lccc}
\toprule
Band (days) & no. of seizures ($n$) & $R$ & Rayleigh $p$ \\
\midrule
0.8--1.2 (circadian) & 29 & 0.553 & $7.5 \times 10^{-5}$ \\
2--5   & 29 & 0.140 & 0.572 \\
3--7   & 29 & 0.066 & 0.885 \\
5--9   & 29 & 0.073 & 0.859 \\
7--14  & 29 & 0.062 & 0.897 \\
10--20 & 29 & 0.094 & 0.776 \\
14--28 & 29 & 0.212 & 0.275 \\
\bottomrule
\end{tabular}
\end{table}

Table~\ref{tab:band_scan} summarises phase-locking statistics across circadian and multi-day bands. Across all tested multi-day ranges (2--28 days), seizure onset phases did not deviate significantly from uniformity. Resultant vector lengths were low (e.g., $R=0.14$ for 2--5 days; $R=0.21$ for 14--28 days), and none reached statistical significance under the Rayleigh test, including after false discovery rate (FDR) correction. These findings provide no robust evidence of multi-day phase locking in this dataset, although modest concentration in the 14--28 day range warrants further investigation in longer or multi-patient recordings.

In contrast, seizures ($n=29$) were non-uniformly distributed over circadian IBI phase (0.8--1.2 days; $R=0.55$, Rayleigh $p=7.5\times10^{-5}$). The Rayleigh test confirmed significant deviation from uniformity ($p=7.5\times10^{-5}$), indicating that seizure onsets were not evenly distributed across the circadian cycle.. The circular mean phase corresponded approximately to the afternoon period, indicating increased seizure occurrence during late-day hours.

\begin{figure}[t]
  \centering
  \includegraphics[width=\columnwidth]{Figures/polar_rose.png}
  \caption{ircadian phase distribution of seizure onsets relative to band-limited IBI circadian phase (0.8--1.2 days) in a single patient ($n=29$ seizures). (a) Polar scatter showing individual seizure phases with mean resultant vector ($R=0.55$). (b) Rose histogram (12 bins) illustrating clustering in late-day hours. The Rayleigh test confirms significant deviation from uniformity ($p=7.5\times10^{-5}$), indicating that seizure onsets preferentially occur at a specific circadian phase rather than being evenly distributed across the cycle.}
  \label{fig:circadian_phase_locking}
\end{figure}

The resultant vector length indicates substantial phase concentration, with seizure onsets clustering around a specific circadian phase (Figure~\ref{fig:circadian_phase_locking}). Although seizure timings were self-reported and may be imprecise, particularly for nocturnal events, the observed concentration suggests a stable circadian modulation in this dataset.

To contextualise these findings, exploratory logistic models were fitted using (i) clock time (time-of-day) and (ii) physiological circadian phase as predictors. Time-of-day reflects external clock time (e.g., hour of day), whereas circadian phase reflects the position within the individual’s underlying biological rhythm estimated from wearable heart activity. These variables are related but not identical, as biological rhythms may shift relative to clock time due to sleep timing or behavioural patterns.
Discriminative performance was modest: time-of-day alone yielded AUC = 0.595, circadian phase alone yielded AUC = 0.603, and sleep score alone yielded AUC = 0.514. These models are not intended as forecasting systems but provide a benchmark indicating that circadian phase captures weak yet detectable structure beyond simple clock-time effects.

\section{Discussion and Future Work}

This single-patient analysis provides preliminary evidence that seizure timing aligns with circadian phase derived from wearable IBI, while multi-day phase locking was not robust in this dataset. These findings suggest that modelling seizure timing purely in clock time may overlook physiologically meaningful oscillatory structure.

Cycle-aware modelling is not intended to replace established seizure forecasting approaches, but to augment them. Many existing time-series and machine-learning models represent temporal dependence using lagged predictor variables or linear autoregressive structures \cite{box_time_2015}. While effective for short-term dynamics, such approaches do not explicitly encode position within an underlying biological rhythm. Incorporating oscillatory phase as an explicit state variable may therefore provide a complementary and biologically grounded representation of temporal dependence.

Prior work has shown that seizure risk may emerge from the interaction of multiple rhythms (e.g., circadian and multi-day cycles). Although multi-day phase locking was not significant in this dataset, modest concentration in the 14--28 day range suggests that longer recordings or multi-patient analysis may reveal multi-scale structure. Future work should examine whether seizure likelihood is highest at specific combinations of circadian and multi-day phase, rather than within a single rhythm alone.

We briefly examined sleep summary measures; however, daily-aggregated sleep scores provide limited temporal resolution for phase-based analysis. Future investigations should consider higher-resolution sleep-derived variables (e.g., sleep timing, REM proportion, sleep fragmentation) and additional wearable signals (e.g., heart rate variability components, activity rhythms, stress proxies). Different physiological systems may express distinct rhythmic profiles, and their interaction may offer richer cycle-aware features for seizure modelling.

From a patient perspective, identifying preferred circadian vulnerability windows may support adaptive self-management strategies, such as activity planning, stress reduction, or medication timing discussions with clinicians. Importantly, phase-aware models could enable interfaces that communicate fluctuating risk states rather than static probabilities, aligning with HCI goals of interpretability and actionable feedback.

Circadian modulation of seizure timing may also have implications for chronotherapy, where medication dosing is aligned with biological rhythms \cite{smolensky_biological_2012}. However, causal inference cannot be established from this analysis, and clinical decisions should not be based on single-patient observational findings.

\paragraph{Limitations} First, seizures were self-reported and may be imprecisely timed or underreported, particularly for nocturnal events. Under-detection of night-time seizures could bias the observed phase distribution toward daytime hours. However, the strength of circadian phase concentration suggests that incomplete nocturnal reporting alone is unlikely to fully account for the observed pattern. Objective seizure detection methods, including wearable accelerometry, heart-rate–based detection, and electrographic monitoring devices, may improve temporal accuracy in future studies. Second, the analysis is restricted to a single patient, limiting generalisability. Third, although zero-phase filtering was used, band-limited phase estimation assumes relatively stable oscillatory structure and may be sensitive to nonstationarity. Alternative approaches such as wavelet-based time–frequency decomposition, adaptive state-space oscillation models, or empirical mode decomposition may better accommodate nonstationary rhythms and will be explored in future work. Finally, phase locking reflects association rather than causation; behavioural schedules, medication timing, or environmental factors may contribute to the observed structure.

\paragraph{Future work.} Future work will extend this proof-of-concept analysis in several directions. First, we will replicate the approach across larger and clinically stratified cohorts to assess generalisability. Second, longer-duration recordings will enable more reliable characterisation of multi-day rhythms and potential multi-scale interactions. Third, incorporating behavioural and medication covariates will help disentangle physiological modulation from external scheduling effects. Finally, we aim to develop and evaluate phase-aware risk models under blocked temporal validation and explore multi-scale cyclical causal modelling across interacting wearable signals.

\bibliographystyle{ACM-Reference-Format}
\bibliography{references}

@article{boashash_estimating_1992,
	title = {Estimating and interpreting the instantaneous frequency of a signal. {I}. {Fundamentals}},
	volume = {80},
	issn = {1558-2256},
	doi = {10.1109/5.135376},
	number = {4},
	urldate = {2026-02-23},
	journal = {Proceedings of the IEEE},
	author = {Boashash, B.},
	year = {1992},
	pages = {520--538},
}

@incollection{smolensky_biological_2012,
	title = {Biological {Rhythms}, {Drug} {Delivery}, and {Chronotherapeutics}},
	isbn = {978-1-4614-0881-9},
	language = {en},
	urldate = {2026-02-26},
	booktitle = {Fundamentals and {Applications} of {Controlled} {Release} {Drug} {Delivery}},
	publisher = {Springer},
	author = {Smolensky, Michael H. and Siegel, Ronald A. and Haus, Erhard and Hermida, Ramon and Portaluppi, Francesco},
	editor = {Siepmann, Juergen and Siegel, Ronald A. and Rathbone, Michael J.},
	year = {2012},
	pages = {359--443},
}

@article{stirling_forecasting_2021,
	title = {Forecasting {Seizure} {Likelihood} {With} {Wearable} {Technology}},
	volume = {12},
	issn = {1664-2295},
	doi = {10.3389/fneur.2021.704060},
	language = {English},
	urldate = {2026-02-18},
	journal = {Frontiers in Neurology},
	author = {Stirling, Rachel E. and Grayden, David B. and D'Souza, Wendyl and Cook, Mark J. and Nurse, Ewan and Freestone, Dean R. and Payne, Daniel E. and Brinkmann, Benjamin H. and Pal Attia, Tal and Viana, Pedro F. and Richardson, Mark P. and Karoly, Philippa J.},
	year = {2021},
	note = {Publisher: Frontiers},
}

@article{quilter_digital_2024,
	title = {A {Digital} {Intervention} for {Capturing} the {Real}-{Time} {Health} {Data} {Needed} for {Epilepsy} {Seizure} {Forecasting}: {Protocol} for a {Formative} {Co}-{Design} and {Usability} {Study} ({The} {ATMOSPHERE} {Study})},
	volume = {13},
	copyright = {Unless stated otherwise, all articles are open-access distributed under the terms of the Creative Commons Attribution License (http://creativecommons.org/licenses/by/2.0/), which permits unrestricted use, distribution, and reproduction in any medium, provided the original work ("first published in JMIR Research Protocols...") is properly cited with original URL and bibliographic citation information. The complete bibliographic information, a link to the original publication on http://www.researchprotocols.org/, as well as this copyright and license information must be included.},
	shorttitle = {A {Digital} {Intervention} for {Capturing} the {Real}-{Time} {Health} {Data} {Needed} for {Epilepsy} {Seizure} {Forecasting}},
	doi = {10.2196/60129},
	language = {EN},
	number = {1},
	urldate = {2026-02-18},
	journal = {JMIR Research Protocols},
	author = {Quilter, Emily E. V. and Downes, Samuel and Deighan, Mairi Therese and Stuart, Liz and Charles, Rosie and Tittensor, Phil and Junges, Leandro and Kissack, Peter and Qureshi, Yasser and Kamaraj, Aravind Kumar and Brigden, Amberly},
	year = {2024},
	pages = {e60129},
}

@article{mason_heart_2024,
	title = {Heart {Rate} {Variability} as a {Tool} for {Seizure} {Prediction}: {A} {Scoping} {Review}},
	volume = {13},
	issn = {2077-0383},
	shorttitle = {Heart {Rate} {Variability} as a {Tool} for {Seizure} {Prediction}},
	doi = {10.3390/jcm13030747},
	number = {3},
	urldate = {2026-02-18},
	journal = {Journal of Clinical Medicine},
	author = {Mason, Federico and Scarabello, Anna and Taruffi, Lisa and Pasini, Elena and Calandra-Buonaura, Giovanna and Vignatelli, Luca and Bisulli, Francesca},
	year = {2024},
	pmid = {38337440},
	pmcid = {PMC10856437},
	pages = {747},
}

@article{karoly_multiday_2021,
	title = {Multiday cycles of heart rate are associated with seizure likelihood: {An} observational cohort study},
	volume = {72},
	issn = {2352-3964},
	shorttitle = {Multiday cycles of heart rate are associated with seizure likelihood},
	doi = {10.1016/j.ebiom.2021.103619},
	urldate = {2026-02-18},
	journal = {EBioMedicine},
	author = {Karoly, Philippa J. and Stirling, Rachel E. and Freestone, Dean R. and Nurse, Ewan S. and Maturana, Matias I. and Halliday, Amy J. and Neal, Andrew and Gregg, Nicholas M. and Brinkmann, Benjamin H. and Richardson, Mark P. and La Gerche, Andre and Grayden, David B. and D'Souza, Wendyl and Cook, Mark J.},
	year = {2021},
	pages = {103619},
}

@article{baud_multi-day_2018,
	title = {Multi-day rhythms modulate seizure risk in epilepsy},
	volume = {9},
	copyright = {2017 The Author(s)},
	issn = {2041-1723},
	doi = {10.1038/s41467-017-02577-y},
	language = {en},
	number = {1},
	urldate = {2026-02-18},
	journal = {Nature Communications},
	author = {Baud, Maxime O. and Kleen, Jonathan K. and Mirro, Emily A. and Andrechak, Jason C. and King-Stephens, David and Chang, Edward F. and Rao, Vikram R.},
	year = {2018},
	pages = {88},
}

@book{box_time_2015,
	title = {Time {Series} {Analysis}: {Forecasting} and {Control}},
	isbn = {978-1-118-67492-5},
	shorttitle = {Time {Series} {Analysis}},
	language = {en},
	publisher = {John Wiley \& Sons},
	author = {Box, George E. P. and Jenkins, Gwilym M. and Reinsel, Gregory C. and Ljung, Greta M.},
	month = may,
	year = {2015},
}

@book{fisher_statistical_1995,
	title = {Statistical {Analysis} of {Circular} {Data}},
	isbn = {978-0-521-56890-6},
	language = {en},
	publisher = {Cambridge University Press},
	author = {Fisher, N. I.},
	month = oct,
	year = {1995},
}

@article{welch_use_1967,
	title = {The use of fast {Fourier} transform for the estimation of power spectra: {A} method based on time averaging over short, modified periodograms},
	volume = {15},
	issn = {1558-2582},
	shorttitle = {The use of fast {Fourier} transform for the estimation of power spectra},
	doi = {10.1109/TAU.1967.1161901},
	number = {2},
	urldate = {2026-02-23},
	journal = {IEEE Transactions on Audio and Electroacoustics},
	author = {Welch, P.},
	year = {1967},
	pages = {70--73},
}

@article{gregg_seizure_2023,
	title = {Seizure occurrence is linked to multiday cycles in diverse physiological signals},
	volume = {64},
	copyright = {© 2023 International League Against Epilepsy.},
	issn = {1528-1167},
	doi = {10.1111/epi.17607},
	language = {en},
	number = {6},
	urldate = {2026-02-18},
	journal = {Epilepsia},
	author = {Gregg, Nicholas M. and Pal Attia, Tal and Nasseri, Mona and Joseph, Boney and Karoly, Philippa and Cui, Jie and Stirling, Rachel E. and Viana, Pedro F. and Richner, Thomas J. and Nurse, Ewan S. and Schulze-Bonhage, Andreas and Cook, Mark J. and Worrell, Gregory A. and Richardson, Mark P. and Freestone, Dean R. and Brinkmann, Benjamin H.},
	year = {2023},
	pages = {1627--1639},
}

@article{butterworth1930theory,
	title = {On the theory of filter amplifiers},
	volume = {7},
	number = {6},
	journal = {Wireless Engineer},
	author = {{Butterworth, Stephen}},
	year = {1930},
	pages = {536--541},
}

\end{document}